\documentclass[aps,pra,twocolumn,superscriptaddress]{revtex4-1}
\usepackage[english]{babel}
\usepackage{amsmath,amsthm,lipsum}
\usepackage{amsfonts}
\usepackage[pdfborder={0 0 0}, colorlinks=true, urlcolor=blue,linkcolor=blue, citecolor=blue]{hyperref}
\usepackage{color}
\usepackage{graphicx}
\usepackage{epstopdf}
\usepackage{float}
\usepackage{subfigure}

\usepackage[normalem]{ulem} %makes underlining possible: \uline \uwave \sout
\usepackage{color}%makes \textcolor{grey}{text} available

\begin{document}
\title{Anomalous wave-packet transport on boundaries of Floquet topological systems}
\author{Xin-Xin Yang}
\affiliation{School of Physics and Key Laboratory of Quantum State Construction and Manipulation (Ministry of Education), Renmin University of China, Beijing 100872, China}

\author{Kai-Ye Shi}
\affiliation{State Key Laboratory for Mesoscopic Physics, School of Physics, Frontiers Science Center for Nano-optoelectronics, $\&$ Collaborative Innovation Center of Quantum Matter, Peking University, Beijing 100871, China}
\affiliation{Hefei National Laboratory, Hefei 230088, China}

\author{F. Nur \"Unal}
\thanks{fnu20@cam.ac.uk}
\affiliation{TCM Group, Cavendish Laboratory,~University of Cambridge, JJ Thomson Avenue, Cambridge CB3 0HE, United Kingdom\looseness-1}
\affiliation{School of Physics and Astronomy,~University of Birmingham, Edgbaston, Birmingham B15 2TT, United Kingdom\looseness-1}

\author{Wei Zhang}
\thanks{wzhangl@ruc.edu.cn}
\affiliation{School of Physics and Key Laboratory of Quantum State Construction and Manipulation (Ministry of Education), Renmin University of China, Beijing 100872, China}
\affiliation{Beijing Academy of Quantum Information Sciences, Beijing 100872, China}
\affiliation{Beijing Key Laboratory of Opto-electronic Functional Materials and Micro-nano Devices, Renmin University of China, Beijing 100872, China}

\date{\today }

\begin{abstract}
A two-dimensional periodically driven (Floquet) system with zero winding number in the absence of time-reversal symmetry is usually considered topologically trivial. Here, we study the dynamics of a Gaussian wave packet placed at the boundary of a two-dimensional driven system with zero winding numbers but multiple valley-protected edge states that can be realized in a square Raman lattice, and investigate the unidirectionally propagating topological edge currents. By carefully tuning the initial parameters of the wave packet including its spin polarization as well as the initial time of the periodic driving, we control the population of different edge states, where the speed of the resulting propagation establishes a direct correspondence with the target dispersions across different gaps and valleys. 
Interestingly, we find that the edge states at different valleys in the $\pi$ gap can hybridize and form bowtie-shaped edge bands fully detached from the bulk. This phase, not only presents a favorable regime with narrower bulk bands, but also exhibits distinct edge dynamics where the majority of particles bounce back-and-forth confined to a boundary while a small portion can follow a chiral transport around the sample.  
\end{abstract}

\maketitle

%%%%%%%%%%%%%
\section{Introduction}
Topologically protected boundary phenomena have gained a tremendous amount of attention in a variety of contexts of physics. One important example is topological insulators, which feature pairs of helical edge modes protected from scattering by time-reversal symmetry~\cite{Hasan2010,Qi2011,Markus2007,Hsieh2007}. These studies have expanded the classification of phase transitions beyond the Landau-Ginzburg-Wilson paradigm~\cite{Po2017,Kruthoff2017,Slager2013}, and triggered the exploration and investigation of topological quantum phases. A gapped topological phase in a free-fermion system is characterized by a bulk topological invariant defined in the ground state, which protects edge states at the system boundary via a relation called bulk-boundary correspondence. For example, non-interacting two-dimensional (2D) systems without additional symmetries are characterized by the Chern number ($C$), which directly determines the net number of one-dimensional (1D) chiral edge states protected by topology. From an experimental perspective, this has been observed as quantized Hall conductance and interpreted with the aid of Chern numbers in condensed matter settings~\cite{Laughlin1981,Halperin1982,Klitzing1980,ferguson2023direct}, while chiral edge states have been detected in photonic crystals~\cite{Hafezi2013,Plotnik2014} and ultracold atomic settings~\cite{StuhlScience15,mancini2015observation,Yao2024,braun2024real}.

The periodic driving technique has been widely applied in experiments for Hamiltonian engineering, which has unlocked novel capabilities such as in settings involving ultracold atoms and optical lattices and enabled the exploration of Floquet topological phenomena~\cite{Cooper2019,EckardtRMP2017,jotzu2014experimental,flaschner2016experimental,Karen2020}.  Through stroboscopic measurements, periodically driven systems can be described by an effective Floquet Hamiltonian exhibiting discrete translational symmetry in the temporal dimension. The eigenenergies of this Floquet Hamiltonian form a quasienergy spectrum in which a $\pi$ gap appears due to the discrete temporal symmetry.  This unique $\pi$ gap in Floquet systems can support topological edge modes, leading to novel Floquet topological phases that have no static counterparts~\cite{KitagawaPRB2010,Rudner2013,Lababidi2014,RoyHarperPRB2017,Rudner2020}. The static topological invariant, e.g., the Chern number, can no longer capture the phase accumulated over a period, and one has to introduce new invariants to fully describe the bulk topological characteristics, such as winding numbers defined for specific gaps~\cite{Rudner2013,Nur2019}, or even for specific points~\cite{Shi2022}. The existence of such topological invariants in bulk spectrum has been witnessed in cold atomic gases by the measurement of transverse deflections~\cite{Karen2020} and band inversion surfaces~\cite{Zhang2023}. Meanwhile, the chiral edge transport has been observed in lattices of photonic waveguides~\cite{Lukas2017,Mukherjee2017}
with first signatures in optical lattices recently appearing~\cite{braun2024real,Miguel2023}. These however mostly remain limited to Floquet topological phases with minimal numbers of edge modes, leaving the interplay of multiple edge states in more complicated settings an open question.

In this paper, we investigate topological transport of edge modes in Floquet topological systems by analyzing the dynamics of Gaussian wave packets initially placed at the boundary of a 2D lattice~\cite{Miguel2023}.
By tuning the initial parameters, such as the shape, momentum and polarization, of a Gaussian wave packet, its evolution under stroboscopic measurements can be predominantly confined to the boundary, thereby forming an edge current. 
More importantly, by measuring the propagation speed of wave packet, one can directly obtain the group velocity of designated edge states in both the $0$ and $\pi$ gaps, allowing for a clear detection of edge state and a quantitative characterization of its dispersion. Combined with techniques for probing the bulk topology such as via band inversion surfaces or measuring the Berry curvature~\cite{Zhang2023,Karen2020}, our scheme enables a full test of bulk-boundary correspondence, a 
fundamental principle of topological states, in periodically driven systems. Further, we find that if edge states with different chiralities coexist in the same gap, these two edge states can hybridize and fully detach from the bulk gap, forming a bowtie-shaped dispersion in the middle of the gap~\cite{PhysRevB.93.245406}. This so-called bowtie-shaped edge band can exhibit exotic dynamic behavior, with two edge states with different chiralities scattering into each other at the corner of the lattice, resulting in a propagation bouncing back and forth at the boundary. 

%%%%%%%%%%%%%

%\section*{Results}
\section{Introduction of Model}
\begin{figure}[t]
    \centering
    \includegraphics[width=1\linewidth]{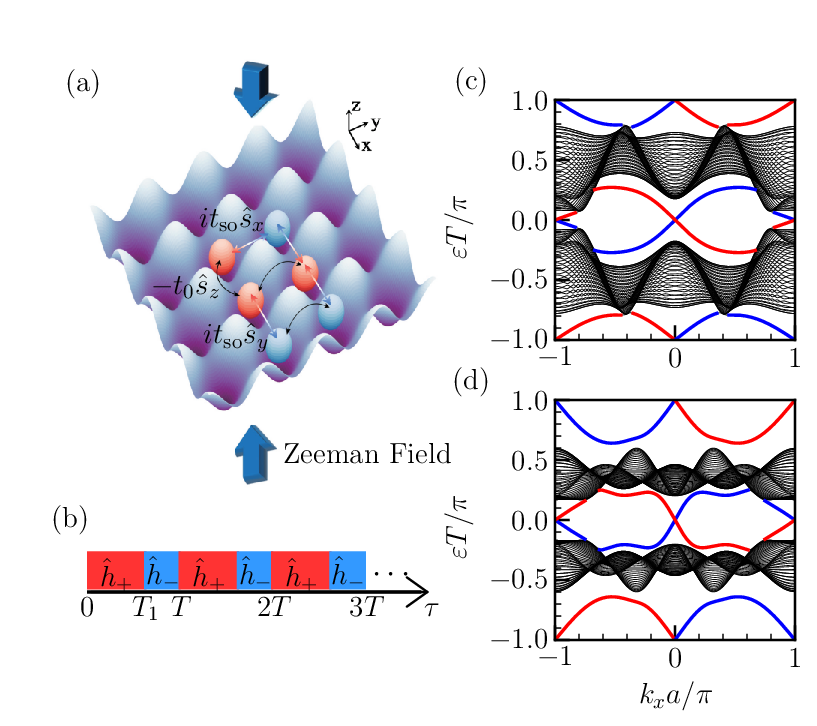}
    \caption{ (a) Illustration of Hamiltonian given in Eq.~(\ref{Hreal}) in a 2D square lattice. (b) Two-stage periodic driving protocol in the form of a step function. (c), (d) Quasienergy spectra of the effective Floquet Hamiltonian on a cylinder periodic along the $\mathbf{x}$ direction and open along the $\mathbf{y}$ direction. Blue and red lines depict the edge states located at the top and bottom boundaries of the cylinder, respectively. The hopping strength with spin flip $t_{\text{so}} = 0.25E_r$ in (c) and $0.5E_r$ in (d) with $E_r$ the recoil energy.  Other parameters are $t_0 =  0.5E_r$, $T_1 = 3/5T$, $T=E_r$, $m_z = 2E_r$ and $N_y = 50$.  }
    \label{setupFig}
\end{figure}

As demonstrated in recent experiments of ultracold $^{87}\text{Rb}$ atomic gases~\cite{Zhang2023,Wu2016}, a 2D square Raman lattice can be engineered by applying two pairs of laser beams with different frequencies (see  Fig.~\ref{setupFig}(a)).
We designate the hyperfine states $|F=1,m_F=-1\rangle$ and $|F=1,m_F=0\rangle$ of $^{87}\text{Rb}$ as the spin-up and spin-down states, and ignore $|F=1,m_F=1\rangle$ state which can be eliminated through large detuning~\cite{Zhang2023,Wu2016}.
The total number of atoms is assumed to be $N_{\text{at}} = 10^5$.
In the tight-binding approximation, the system can be described by a quantum anomalous Hall effect (QAHE) model. The Hamiltonian is given by 
\begin{eqnarray} \label{Hreal}
    \hat{H} = \sum_{\vec{r}} m_z \hat{s}_z\hat{c}^{\dagger}_{\vec{r}}\hat{c}_{\vec{r}} -\sum_{\vec{r},\vec{d}} (t_{0} \hat{s}_z - i t_{\text{so}} \hat{s}_{\vec{d}}) \hat{c}^{\dagger}_{\vec{r}}\hat{c}_{\vec{r}+\vec{d}} +H.c.,  
\end{eqnarray}
where $\hat{c}^{\dagger}_{\vec{r}}$ and $\hat{c}_{\vec{r}}$ are the creation and annihilation operators of particles at lattice site $\vec{r}$, $t_{\text{so}}$($t_0$) represents the hopping strength with (without) spin flip, and $m_z = \delta/2$ is the longitudinal Zeeman field with detuning $\delta$. For convenience, we set $\hbar=1$. The nearest-neighbor vectors $\vec{d}_1 = (a,0)$ and $\vec{d}_2= (0,a)$ with lattice constant $a$ correspond to hopping along the $\mathbf{x}$ and $\mathbf{y}$ directions, respectively. The Pauli operator $\hat{s}_{\vec{d}}$ in the spin space is $\hat{s}_y$ for $\vec{d} =\vec{d}_1$, and takes $\hat{s}_x$ for $\vec{d} =\vec{d}_2$. We consider an isolated square lattice with $N_x$ and $N_y$ denoting the number of sites in the $\mathbf{x}$ and $\mathbf{y}$ directions respectively. This static Hamiltonian can be expressed in Bloch form with respect to quasi-momentum $\mathbf k$ as 
\begin{eqnarray}
	\hat{H}(\mathbf k)=\mathbf h(\mathbf k)\cdot \mathbf{s},
\end{eqnarray}
where $\mathbf{h}(\mathbf{k}) = (2t_{\text{so}} \sin k_y a, 2t_{\text{so}} \sin k_x a, m_z - 2t_0 (\cos k_x a + \cos k_y a))$ and $\mathbf{s} = (\hat{s}_x, \hat{s}_y, \hat{s}_z)$. 

Analyzing the symmetries of this model will be useful to understand the topological properties of the system.
In a square lattice, there exist high symmetry points: the $\Gamma$ point with $\mathbf{k} = (0, 0)$, the $M$ point with $\mathbf{k} = (\pi/a, \pi/a)$, the $X$ point with $\mathbf{k} = (\pi/a, 0)$, and the $Y$ point with $\mathbf{k} = (0, \pi/a)$ which maps onto $X$ under $C_4$ symmetry. Correspondingly, we obtain $\mathbf{h}(\Gamma) = (0, 0, m_z - 4t_0)$, $\mathbf{h}(M) = (0, 0, m_z + 4t_0)$ and $\mathbf{h}(X) = (0, 0, m_z)$,
with $\hat{H}(k)$ satisfying $C_4$ (and hence also parity) symmetry $(k_x, k_y; \hat{s}_x, \hat{s}_y) \rightarrow (k_y, -k_x; -\hat{s}_y, \hat{s}_x)$~\cite{Shi2022}.

To induce an anomalous Floquet topological phase, we consider a time-dependent Hamiltonian $\hat{H}(\mathbf{k}, t)$ with a two-stage periodic driving protocol, as illustrated in Fig.~\ref{setupFig}(b). For simplicity, we assume that in the first stage  ($0 \leq \tau < T_1$) of a driving period $T$, the Zeeman field takes a constant value $m_z$, while in the second stage ($T_1 \leq \tau < T$), the Zeeman field is reversed to $-m_z$. This can be achieved by periodically reversing the detuning $\delta$. For convenience, we denote the first (static)
Hamiltonian with $m_z$ as $\hat{h}_{+}$, and the second  
Hamiltonian with $-m_z$ as $\hat{h}_{-}$. Starting from the three high symmetry points, we can readily obtain $\hat{h}_{+}(\Gamma) = -\hat{h}_{-}(M)$, $\hat{h}_{+}(M) = -\hat{h}_{-}(\Gamma)$, and $\hat{h}_{+}(X) = -\hat{h}_{-}(X)$.  
Importantly, the periodic driving protocol illustrated in Fig.~\ref{setupFig}(b) preserves the $C_4$ symmetry, which is manifested in the quasienergy spectrum. This symmetry enables the emergence of exotic edge states, which will be explored in detail in the following sections. The dynamical characteristics of this periodically driven system after  
each complete period can be described by an effective Floquet Hamiltonian
\begin{eqnarray}
    \hat{H}^{\tau_0}_F(\mathbf k)= i\frac{\log U^{\tau_0}_T}{T},
\end{eqnarray}
where  $\hat{U}^{\tau_0}_T = \hat{\mathcal{T}} e^{-i\int_{\tau_0}^{T+\tau_0} \hat{H}(\mathbf{k}, \tau) dt}$ is the evolution operator over one period, with $\hat{\mathcal{T}}$ representing the time-ordering operator. Here, $\tau_0$ defines different starting times, referred to as different Floquet gauges. 
The Floquet quasienergy spectra is defined as $\hat{H}^{\tau_0}_F(\mathbf{k}) |\psi_m (\mathbf{k})\rangle^{\tau_0} = \varepsilon_m(\mathbf{k}) |\psi_m (\mathbf{k})\rangle^{\tau_0}$, where $m = 0, 1$ is the band index. We focus on the first Floquet Brillouin zone (FBZ), within which the quasienergy spectrum is restricted to the range $(-\pi/T, \pi/T]$. Notably, the Floquet quasienergy spectrum for different FBZs exhibits periodicity, repeating with a period of $2\pi/T$. We can define a gauge transformation operator $\mathcal{B}(\tau_1, \tau_0)$ to change the effective Floquet Hamiltonian from initial time $\tau_0$ to $\tau_1$, i.e.,
\begin{eqnarray}
    \hat{H}^{\tau_1}_F(\mathbf k) = \mathcal{B}(\tau_1, \tau_0)\hat{H}^{\tau_0}_F(\mathbf k) \mathcal{B}^{-1}(\tau_1, \tau_0).
\end{eqnarray}
We can readily obtain the corresponding eigenvalue equation as
\begin{eqnarray}
    \hat{H}^{\tau_1}_F(\mathbf k) |\psi_m (\mathbf{k})\rangle^{\tau_1} = \varepsilon_m(\mathbf{k}) |\psi_m (\mathbf{k})\rangle^{\tau_1},
\end{eqnarray}
where
\begin{eqnarray} 
    |\psi_m (\mathbf{k})\rangle^{\tau_1} \equiv \mathcal{B}(\tau_1, \tau_0)|\psi_m (\mathbf{k})\rangle^{\tau_0}.
\end{eqnarray}
Therefore, choosing different Floquet gauges, i.e., different starting times, results in the same Floquet quasienergy spectra but with different corresponding eigenstates. Due to the self-replicating nature between different FBZs, edge states can exist not only in the gap of $0$ quasienergy (referred to as the $0$ gap) but also in the gaps of $\pm \pi/T$ quasienergy (referred to as the $\pi$ gap). Thus, the Chern number ($C$) alone is insufficient to determine the number and chirality of edge states. Winding numbers related to the $0$ gap ($W_0$) and $\pi$ gap ($W_{\pi}$) should be introduced to help classify different topological phases~\cite{Rudner2013}.

Bands crossings occur at $\Gamma$ and $M$ points in our system under periodic driving, followed by creation of edge states at relevant momenta. The interplay between $C_4$ symmetry and charge conjugation symmetry, ensures that the edge states at the $\Gamma$ and $M$ points possess opposite chiralities~\cite{Shi2022}.The winding number for the corresponding gap increases by 1 whenever a band crossing takes place at $\Gamma$ and decreases by 1 for the ones at $M$.  Therefore, when edge states exist at both the $\Gamma$ and $M$ points in a given gap, the winding number will no longer match the number of edge states. Ref.~\cite{Shi2022} identified various different topological phases where Chern numbers and winding numbers $(C, W_0, W_{\pi})$ indeed fail to capture the edge states; e.g.~a phase with $(C, W_0, W_{\pi})=(1,1,0)$ can harbor no edge states or two pairs at different valleys with opposite chiralities in the $\pi$ gap, while the unit winding number in the $0$ gap could arise from a single pair of edge states, three pairs of edge states or even more. Neither the Chern number nor the winding numbers alone are sufficient to distinguish different topological phases or characterize the number of edge modes in systems with valley symmetry. To address this limitation, new topological invariants, $\nu_{F,0/\pi}^\Gamma$ and $\nu_{F,0/\pi}^M$, have been proposed to describe the edge states around the high symmetry points $\Gamma$ and $M$, respectively~\cite{Shi2022}. These invariants provide a robust framework for establishing the bulk-edge correspondence in a valley-resolved manner. As an illustrative example, Fig.~\ref{setupFig}(c) presents a case with $(C, W_0, W_{\pi}) = (0,0,0)$, which, based on traditional Chern and winding number analyses, would be classified as topologically trivial. However, the system clearly exhibits four edge states in both the  $0$ gap and $\pi$ gap, occurring at $k_x = 0$ and $k_x = \pi/a$, respectively. The corresponding topological invariants, $(\nu_{F,0}^\Gamma, \nu_{F,\pi}^\Gamma, \nu_{F,0}^M, \nu_{F,\pi}^M)=(1,1,1,1)$, successfully account for these observed edge states, demonstrating the effectiveness of these invariants in capturing the topological features of the system. 

Furthermore, the edge states anchored at different valleys within the $\pi$ gap in Fig.~\ref{setupFig}(c) do not hybridize and remain continuously connected to the bulk bands.  As $t_{\text{so}}$ increases, the bulk band narrows, leading to reduced overlap between the edge and bulk states. Notably, when $t_{\text{so}}$ exceeds a certain threshold, the edge states in the $\pi$ gap fully hybridize, forming midgap bowtie-shaped bands that are completely detached from the bulk as shown in Fig.~\ref{setupFig}(d), while 
the topological invariants $(\nu_{0F}^{\Gamma}, \nu_{F,\pi}^{\Gamma}, \nu_{0F}^M, \nu_{F,\pi}^M)$ remain unchanged. However, this hybridization introduces nontrivial effects on the dynamics of the edge states, as demonstrated below. As the corresponding edge band widens, the group velocity of the edge state increases significantly. For instance, in Fig.~\ref{setupFig}(c), where $t_{\text{so}} = 0.25E_r$ ($E_r$ being the recoil energy), the group velocity of the $0$ gap edge state at $k_x = 0$ is $a/T$, and that of the $\pi$ gap edge state at $k_x = \pi/a$ is $0.66a/T$. By contrast, when $t_{\text{so}} = 0.5E_r$, as shown in Fig.~\ref{setupFig}(d), the group velocity of the $0$ gap edge state at $k_x = 0$ increases to $1.8a/T$, while the $\pi$ gap edge state at $k_x = \pi/a$ reaches $1.21a/T$.  With narrower bulk bands, this regime with larger $t_{\text{so}}$ values offers a promising candidate to study edge dynamics where the bowtie-shaped edge bands bring more exotic features into play.

\section{Edge state dynamics in cylindrical geometry}
Using the topological invariants $\nu_{F,0/\pi}^\Gamma$ and $\nu_{F,0/\pi}^M$, one can theoretically identify the edge states individually. Experimentally, bulk topological invariants in the QAHE model have been successfully determined by detecting the band inversion surfaces~\cite{Zhang2023} under periodic driving and by measuring the topological charges of band touching points in a gap-resolved way~\cite{Karen2020,Nur2019}. The observation of Floquet topological edge states remains limited, which mostly concentrates on transport along the edge in the presence of a single edge channel overall or maximum one per gap~\cite{Lukas2017,Mukherjee2017,braun2024real,Miguel2023}. Here, we investigate topological edge transport in a broader class of parameters, with valley topological invariants, larger winding numbers and multiple edge states, by also analyzing the propagation velocity of edge currents which we employ to assess topological edge states.

We first consider a cylindrical geometry and, without loss of generality, imagine a 2D square lattice with periodic boundary condition along the $\mathbf{x}$ direction and open boundary condition in the $\mathbf{y}$ direction. In this configuration, edge states propagate along either the top or bottom boundary of the cylinder, as indicated by blue and red lines in Figs.~\ref{setupFig}(c) and (d) respectively, with the corresponding dispersion relation defined by $k_x$. In our periodically-driven model due to the underlying symmetries, we observe that the edge modes with $k_x = 0$ always propagate to the right at the top boundary (in the positive $\mathbf{x}$ direction), while the ones at $k_x = \pi/a$ move to the left.
These opposite chiralities play a critical role in distinguishing the edge states associated with the $\Gamma$ or $M$ points, as will be discussed in the subsequent.

We prepare a Gaussian wave packet as the initial state in real space and allow it to evolve under the periodically driven Hamiltonian illustrated in Fig.~\ref{setupFig}(b), following an approach similar to Ref.~\cite{Miguel2023}. The initial state $|{\Psi(0)}\rangle$ is expressed in real space using the basis set $\{|{x,y,\uparrow}\rangle,|{x,y,\downarrow}\rangle\}$ as,
\begin{eqnarray}
	\langle{x,y,\uparrow/\downarrow|\Psi(0)}\rangle = \frac{\mathcal{C}_{\uparrow/\downarrow}}{\mathcal{N}}e^{-\frac{(x-x_0)^2}{4\sigma_x^2} -\frac{(y-y_0)^2}{4\sigma_y^2} +iq_xx+iq_yy},
\end{eqnarray}
where $(x_0, y_0)$ signify the mean coordinates, $(\sigma_x, \sigma_y)$ denote the widths of the wave packet, and $\mathcal{N}$ is the normalization factor. The initial polarization of the wave packet in spin space is determined by the vector $(\mathcal{C}_{\uparrow},\mathcal{C}_{\downarrow} )^T$, which provides an extra knob to tune the overlap with edge states in our system. An initial kick with momentum $(q_x, q_y)$ is allowed to adjust the mean position of the Gaussian wave packet in momentum space. The extent of the wave packet in real and momentum spaces is inversely proportional. Therefore, a sufficiently wide wave packet in real space becomes narrow in momentum space, allowing targeting a specific edge state better.

\begin{figure}[t]
    \includegraphics[width=1\linewidth]{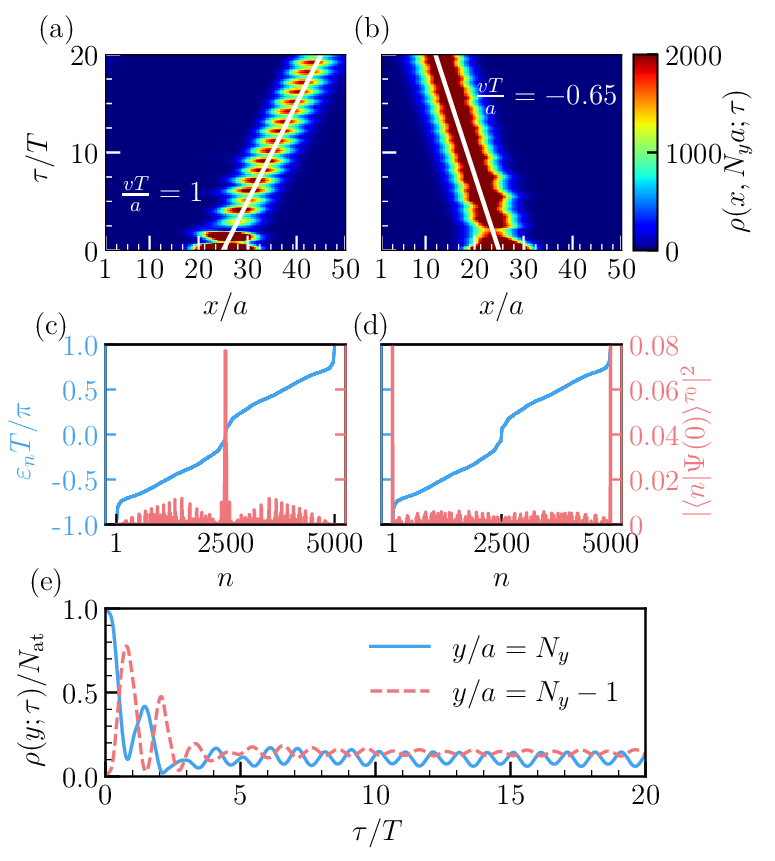}
    \caption{(a) and (b) show the time-dependent density distribution $\rho(x,y;\tau)$ of the wave packet at the top boundary of the cylinder, with $N_x = N_y = 50$. (a) The initial Gaussian wave packet is polarized in the $\mathbf{x}$ direction with $(\mathcal{C}_{\uparrow},\mathcal{C}_{\downarrow}) = (1,1)/\sqrt{2}$, and released without an initial kick $(q_x,q_y)=(0,0)$. (b) The initial wave packet is polarized in the -$\mathbf{x}$ direction with $(\mathcal{C}_{\uparrow},\mathcal{C}_{\downarrow}) = (1,-1)/\sqrt{2}$, and the initial kick is $(q_x,q_y)=(\pi/a,0)$, with remaining parameters 
    $(x_0,y_0) = (N_x/2, N_y)a$ and $(\sigma_x, \sigma_y)=(3,0.1)a$. The period-driven Hamiltonian is consistent with Fig.~\ref{setupFig}(c), where we also tune the Floquet gauge (a) $\tau_0 = 0.68T$ and (b) $\tau_0 = 0$ to increasingly populate the target edge states. The corresponding overlap $|\langle{n|\Psi(0)}\rangle^{\tau_0}|^2$ of the initial wave packet, related to (a) and (b), with the eigenstates of the Floquet Hamiltonian is shown in red in (c) and (d), with blue points representing the corresponding Floquet energies $\varepsilon_n$.  (e) demonstrates the density distribution in the topmost (blue solid line) and sub-top (red dotted line) layers as a function of time.}
    \label{evoPXOY}
\end{figure}

The wave-packet dynamics satisfy the Schr\"{o}dinger equation, and after one complete period, the wave function evolves to $|{\Psi(T)}\rangle^{0}=  e^{-i\hat{h}_-(T-T_1)}e^{-i\hat{h}_+T_1}|{\Psi(0)}\rangle$ under the Floquet gauge $\tau_0 = 0$. This final state, $|{\Psi(T)}\rangle^{0}$, is consistent with stroboscopic measurements, such that $|{\Psi(T)}\rangle^{0} = e^{-i\hat{H}^{0}_F T}|{\Psi(0)}\rangle$. In real space, we denote the eigenstates and eigenenergies of $\hat{H}_F^0$ by $|{n}\rangle^{0}$ and $\varepsilon_n$, respectively. Then, the dynamics over one period can be expanded into 
\begin{eqnarray}
    |{\Psi(T)}\rangle^{0} = \sum_n e^{-i\varepsilon_nT}  \langle n|\Psi(0) \rangle^{0} |{n}\rangle^{0}.
\end{eqnarray}
Thus, the stroboscopic measurement reflects the accumulation of a phase $e^{-i\varepsilon_nT}$ on each Floquet eigenstate $|{n}\rangle^{0}$, with probability determined by $|\langle{n|\Psi(0)}\rangle^{0}|^2$. Increasing the probability of the target edge state while minimizing the contributions of non-target states enhances the accuracy of capturing the edge state dynamics. For instance, in Fig.~\ref{evoPXOY}(d), we aim to observe the $\pi$ gap edge state at $k_x=\pi/a$ along the top boundary of a cylinder. To achieve this, we set $(q_x, q_y) = (\pi/a, 0)$ and $y_0 = N_y a$ with a sufficiently narrow wave packet in the $\mathbf{y}$-direction and broad in the $\mathbf{x}$-direction $(\sigma_x, \sigma_y) = (3, 0.1)a$. Furthermore, the spin polarization is optimized as $(\mathcal{C}_{\uparrow}, \mathcal{C}_{\downarrow}) = (1,-1)/\sqrt{2}$ to increase the overlap with edge states around $\varepsilon_n =\pm \pi/T$, ensuring their probabilities significantly exceed those of other states  [see Fig.~\ref{evoPXOY}(d)]. We numerically calculate the dynamics of the Gaussian wave packet and plot particle density distribution $\rho(x,y=N_ya;\tau) = N_{\text{at}}(|\langle{x,N_ya;\uparrow|\Psi(\tau)}\rangle|^2+|\langle{x,N_ya;\downarrow|\Psi(\tau)}\rangle|^2)$ along the top boundary in Fig.~\ref{evoPXOY}(b). A portion of the Gaussian wave packet remains confined to the top boundary, exhibiting a unidirectional propagating current, clearly demonstrating the presence of the edge state near $k_x = \pi/a$. 

Adjusting the parameters of the initial Gaussian wave packet may  not always yield the desired probability distribution $|\langle{n|\Psi(0)}^{\tau_0}\rangle|^2$ easily. Here, we demonstrate that the Floquet gauge offers an additional tunable parameter to modify the probability distribution~\cite{Miguel2023}. While the energy spectrum of $\hat{H}_F^{\tau_0}$ remains unchanged across different Floquet gauges, the corresponding eigenstates vary, naturally leading to different probability distributions dependent on  $\tau_0$. To explicitly illustrate this, we assume $0 < \tau_0 < T_1$ without loss of generality and express the evolution of the wave packet over $p$ periods as
\begin{eqnarray}
|{\Psi(pT)}\rangle^{\tau_0} =  e^{-i\hat{h}_+\tau_0}\left(\hat{U}_T^{0}\right)^p e^{i\hat{h}_+\tau_0}|{\Psi(0)}\rangle.    
\end{eqnarray}
The difference between $|\Psi(pT - \tau_0)\rangle^{\tau_0}$ and $|\Psi(pT)\rangle^{0}$ can be interpreted as the initial state undergoing a transformation by $e^{i\hat{h}_+\tau_0}$.  Consequently, the overlaps of $\langle{n|\Psi(0)}\rangle^{0}$ and  $\langle{n|e^{i\hat{h}_+\tau_0}|\Psi(0)}\rangle^{0}$ are distinct, as 
$e^{-i\hat{h}_+\tau_0}|{n}\rangle^{0} = |{n}\rangle^{\tau_0}$. In Fig.~\ref{evoPXOY}(c), we plot the probability distribution of $|\langle n|\Psi(0)\rangle|^{\tau_0}$ for $\tau_0=0.68T$. The results show that tuning the Floquet gauge significantly enhances the probability of the $\varepsilon_n=0$,  while simultaneously suppressing the population of $\pi$ gap states. This provides an experimental knob for selectively controlling edge state dynamics.
We show the corresponding dynamics of the wave packet at the top boundary in Fig.~\ref{evoPXOY}(a), where the edge current propagates unidirectionally to the right, consistent with the characteristics of the targeted $0$ gap edge states.

\begin{figure*}[t]
    \includegraphics[width=1\linewidth]{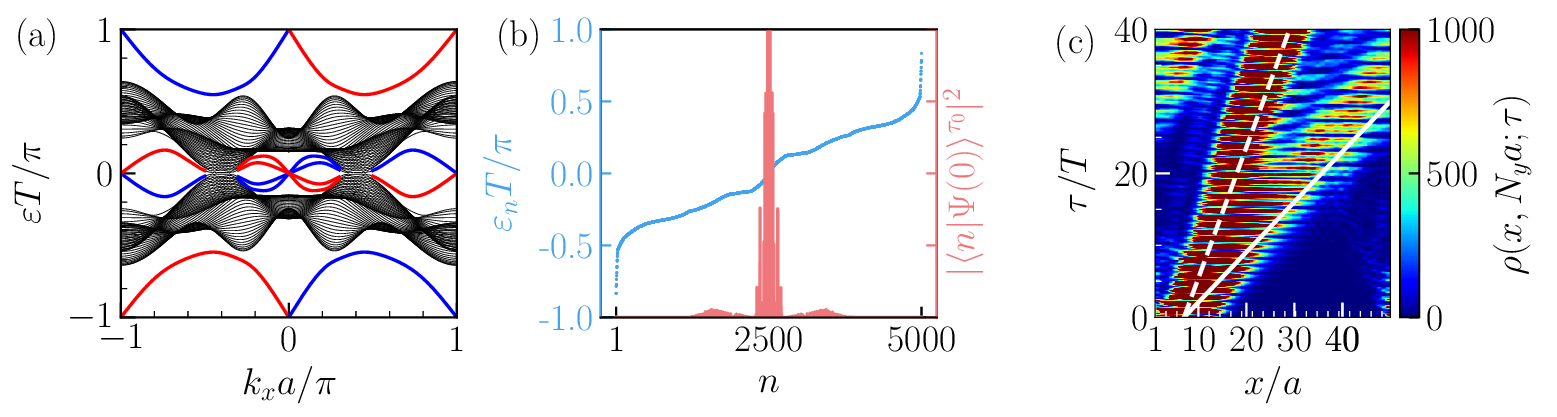}
    \caption{ (a) The quasienergy spectrum of the effective Hamiltonian on a cylinder, with valley topological invariants $(\nu_{F,0}^{\Gamma}, \nu_{F,\pi}^{\Gamma}, \nu_{F,0}^{M}, \nu_{F,\pi}^{M}) = (2, 1, 1, 1)$. The initial Gaussian wave packet is polarized in the $\mathbf{y}$-direction, with $(\mathcal{C}_{\uparrow}, \mathcal{C}_{\downarrow}) = (1, i)/\sqrt{2}$, and initialized without a kick $(q_x, q_y) = (0, 0)$. To enhance the occupation of the target edge state at the $0$ gap, we tune the Floquet gauge to $\tau_0 = 0.6T$. Additional parameters for the wave packet are $(x_0, y_0) = (7, N_y)a$ and $(\sigma_x, \sigma_y) = (3, 0.1)a$.
    (b) The overlap $|\langle n | \Psi(0) \rangle^{\tau_0}|^2$ between this initial wave packet and the eigenstates of the Floquet Hamiltonian is shown in red, while the corresponding Floquet quasienergies $\varepsilon_n$ are depicted as blue points.
    (c) The time-dependent density distribution $\rho(x, y; \tau)$ of the wave packet along the top boundary of the cylinder clearly demonstrates two branches.  The solid white line corresponds to a velocity of $1.43a/T$,  whereas the dashed white line represents a velocity of $0.55a/T$. Other parameters are $t_0 = 0.5E_r$, $T_1 = 3/5T$, $T = 1.2E_r$, $m_z = 2E_r$, and $N_x = N_y = 50$.
    } 
    \label{2111PXOY}
\end{figure*}

Moreover, as depicted in Figs.~\ref{evoPXOY}(a) and (b), the propagation velocity $v$ (white lines) precisely matches the group velocity extracted from the energy spectra.  Further analysis across different parameter regimes and topological numbers consistently confirms that the wave packet propagation velocity aligns with the group velocity. It is important to clarify that the propagation velocity here refers to the time-averaged velocity over an extended period of evolution spanning at least several periods, rather than the micromotion velocity. From the perspective of micromotion, the wave packet at the edge does not remain confined strictly to $y=N_ya$. Instead, it oscillates within several adjacent layers, as shown in Fig.~\ref{evoPXOY}(e).

Our square-Raman-lattice model allows for reaching a range of Floquet topological phases with multiple edge states harbored at different energies and/or valleys. We demonstrate a distinct phase 
characterized by $(\nu_{F,0}^{\Gamma},\nu_{F,\pi}^{\Gamma},\nu_{F,0}^{M},\nu_{F,\pi}^{M}) = (2,1,1,1)$ in Fig.~\ref{2111PXOY}(a), where two edge states are clearly visible at the $\Gamma$ point in $0$ gap.  

According to the previously outlined scheme, the edge state(s) associated with the topological number $\nu_{F,0}^{\Gamma}$ reaches maximum occupation when a Gaussian wave packet is initialized without a kick $(q_x, q_y) = (0,0)$.  We tune the spin polarization of the wave packet to be  aligned along the $\mathbf{y}$-direction and the Floquet gauge to $\tau_0=0.6T$, which are chosen to suppress the bulk state occupation, as shown in Fig.~\ref{2111PXOY}(b). However, since $\nu_{F,0}^{\Gamma} = 2$, the two edge states overlap, making them indistinguishable in Fig.~\ref{2111PXOY}(b). Nevertheless, their distinct group velocities enable separation through wave packet dynamics. After more than 20 driving periods, two clearly distinguishable wave packets emerge along the boundary, as illustrated in Fig.~\ref{2111PXOY}(c). The group velocities of these edge states are represented by the white solid and dashed lines, corresponding to $1.43a/T$ and $0.55a/T$, respectively. Once separated, the wave packets propagate along trajectories that closely follow the solid and dashed lines predicted by the energy spectrum, highlighting the difference in group velocities between the two edge states. We note that here we have chosen an example where the bulk bands are actually touching in the $0$ gap. However, the edge states are still visible locally around each valley which can be accessed in the wave packet dynamics distinctly.

\begin{figure*}[t]
    \includegraphics[width=1.1\linewidth]{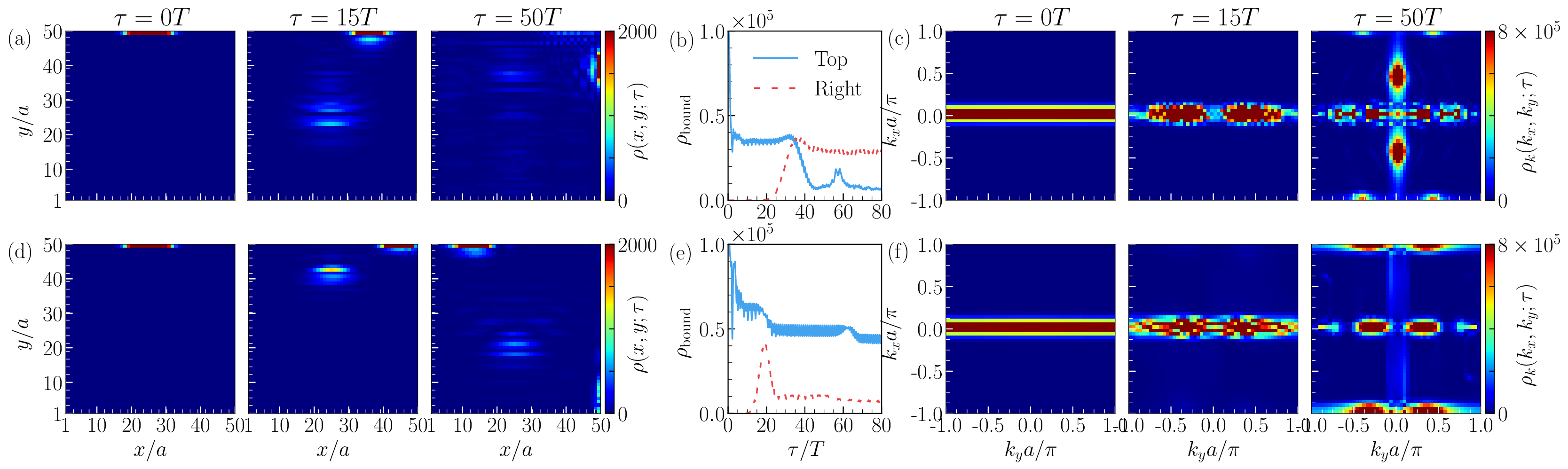}
    \caption{(a) and (d) illustrate the density distribution of wave packets in a square geometry under open boundary condition with $N_x=N_y=50$, at $\tau = 0T$, $15T$, and $50T$. (b) and (e) present the time evolution of the total probability of the wave packets within two layers at the top (blue solid line) and right (red dashed line) boundaries. (c) and (f) depict the corresponding density distributions of the wave functions in momentum space. The first and second rows correspond to the period-driven Hamiltonian parameters of Figs. \ref{setupFig}(c) and (d), respectively. The latter features midgap bowtie-shaped bands fully detached from the bulk, resulting in a considerable portion of the particles bouncing back at the corner as also captured in (f). All other initial parameters are identical, and the initial Gaussian wave packet is characterized by $(\mathcal{C}_{\uparrow},\mathcal{C}_{\downarrow}) = (1,1)/\sqrt{2}$, $(q_x,q_y) = (0,0)$, $(x_0,y_0) = (N_x/2, N_y)a$, and $(\sigma_x, \sigma_y) = (3, 0.1)a$, with the Floquet gauge set to $\tau_0 = 0$.} 
    \label{evoOXOY}
\end{figure*}

\section{Dynamics in a square geometry and effects of bowtie-shaped edge states }
We now consider a square-shaped sample with open boundary condition in both $\mathbf{x}$ and $\mathbf{y}$ directions. 
In 2D systems, the nature of the chiral edge currents is to preserve a definite chirality even after scattering at corners and keep circling around the sample~\cite{Lukas2017,Mukherjee2017}. 
In Fig.~\ref{evoOXOY}(a), we provide an example where the initial Gaussian wave packet predominantly populates the $\pi$ gap states corresponding to the phase shown in Figs.~\ref{setupFig}(c), without an initial kick $(q_x, q_y)=(0,0)$. It is evident that edge current originates from the top boundary and propagates to the right boundary after scattering at the corner. For clarity, we define $\rho_{\text{bound}}$ to denote the total probability of the wave packet within two layers of the boundary as 
\begin{eqnarray}
\rho_{\text{bound}} =  \begin{cases}
                           \sum_{i=0}^1\sum_{s_z} N_{\text{at}}|\langle{y=(N_y-i)a;s_z|\Psi(\tau)}\rangle|^2  \\
                                       \qquad\qquad\qquad\qquad\qquad\qquad\qquad\quad,\text{Top}\\
                           \\
                            \sum_{i=0}^1\sum_{s_z} N_{\text{at}}|\langle{x=(N_x-i)a;s_z|\Psi(\tau)}\rangle|^2 \\
                                        \qquad\qquad\qquad\qquad\qquad\qquad\qquad\quad, \text{Right}
                    \end{cases}   
\end{eqnarray}
for top  and right boundaries, respectively. We present the corresponding $\rho_{\text{bound}}$ in Fig.~\ref{evoOXOY} (b), where the solid blue (dashed red) line depicts the probability at the top (right) boundary. The rapid saturation of the probability at the top boundary, reaching the first plateau, corresponds to unidirectional propagation of the edge current along the top boundary visible in the dynamics. As the edge current reaches the corner, the density at the top boundary depletes, while the probability at the right boundary increases and stabilizes at a new plateau.  This behavior reveals that while a small portion of the edge current gets reflected back upon scattering at the corner, the majority continues traveling to the right boundary and ultimately circulates around the sample, maintaining its original chirality.

Notably, we find that the edge current does not preserve its initial chirality after scattering at the corner in the presence of midgap bowtie-shaped bands, as illustrated in Fig.~\ref{setupFig}(d).  To demonstrate this, we prepare the initial state at the top boundary and show its evolution in Fig.~\ref{evoOXOY}(d), with the corresponding $\rho_{\text{bound}}$ given in Fig.~\ref{evoOXOY}(e). Unlike the behavior observed in Figs.~\ref{evoOXOY}(a) and (b), only a small portion of the edge current transitions to the right boundary after reaching the corner. Instead, the majority reverses direction and continues propagating along the top boundary.  The peak of the dashed red line in Fig.~\ref{evoOXOY}(e) reflects the moment when the edge current reaches the corner, after which we observe that the density at the top boundary attains a plateau. Importantly, we emphasize that this behavior is independent of the boundary shape or the presence of the corner itself. Modifying the boundary on which the initial wave packet resides or altering the geometry of the corner does not influence this characteristic behavior.

To better understand the scattering of the edge current at the corner, we analyze the probability distribution in momentum space, as displayed in Figs.\ref{evoOXOY}(c) and (f). The wave packet is initially centered near $k_x=0$ at $\tau=0$. Before reaching the corner, regardless of whether the edge states are connected to bulk bands, the wave packet remains concentrated around $k_x=0$ as expected. After reaching the corner, however, the scattering characteristics diverge depending on the edge spectra. 
At $\tau=50T$, the wave function predominantly scatters to $k_y=0$ in Fig.\ref{evoOXOY}(c). In contrast, when the edge states are detached from the bulk, the scattering occurs predominantly within the edge band itself, shifting from $k_x=0$ to $k_x=\pm \pi/a$ [Fig.\ref{evoOXOY}(f)]. This observation suggests that when the band is connected, the edge current scatters through the bulk band to another boundary. Conversely, when the band is separated, the edge current is confined to the edge band and scatters internally, transitioning from $k_x=0$ to $k_x=\pi/a$. 	

\section{Summary}
In conclusion, we have investigated the wave packet dynamics in a 2D quantum anomalous Hall effect(QAHE) model that can be experimentally realized by using ultracold atoms in a Raman lattice which allows for reaching a rich phase diagram. By carefully tuning the parameters of the Gaussian wave packet and the Floquet gauge, we have demonstrated that the topological edge transport, corresponding to the valley invariants $\nu^{\Gamma}_{F,0}$, $\nu^{\Gamma}_{F,\pi}$, $\nu^{M}_{F,0}$ and $\nu^{M}_{F,\pi}$ one by one, can be observed. Our analysis shows that after approximately $5$ periods, the propagation velocity of the edge current stabilizes and matches the group velocity, offering a robust method for verifying the bulk-boundary correspondence.  Additionally, we observe the anomalous dynamics in the $\pi$ gap, where edge-state hybridization between different valleys with opposite chiralities leads to edge currents that oscillate back and forth along a single boundary, highlighting unique features of topological behavior in this system.	Our analysis demonstrates the rich interplay of topological edge transport in periodically driven systems and valley-protected phenomena, which is importantly can be probed in experiments with recent advances in local accessibility and control in ultracold atoms.

\section{acknowledgements}
This work is supported by the National Natural Science Foundation of China (Grant Nos. 12074428, 92265208), the National Key R\&D Program (Grant No. 2022YFA1405300).
F.N.\"U.~acknowledges support from the Royal Society via URF/R1/241667, Simons Investigator Award [Grant No.~511029] and Trinity College Cambridge.

\end{document}